\documentclass[12pt,preprint]{aastex}
\usepackage{natbib}
\def\msun{{\rm ~M}_{\odot}}
\def\rsun{{\rm ~R}_{\odot}}
\usepackage[usenames,dvips]{color}
\usepackage{graphicx}

\slugcomment{Accepted for publication in The Astrophysical Journal}
\shorttitle{SYNTHETIC SPECTRA OF AM CVn DONOR STAR}
\shortauthors{SENGUPTA AND TAAM }
\received{2011 May 16}
\accepted{2011 July 06}
\begin{document}

\title{THEORETICAL SPECTRA OF THE AM CVN BINARY SYSTEM
SDSS J0926+3624 : EFFECTS OF IRRADIATION ONTO THE DONOR STAR} 

\author{Sujan Sengupta\altaffilmark{1,2}, and Ronald E. Taam\altaffilmark{2,3}}
\altaffiltext{1}{Indian Institute of Astrophysics, Koramangala 2nd Block, Bangalore 560034, India; sujan@iiap.res.in}
\altaffiltext{2}{Academia Sinica Institute of Astronomy and Astrophysics - TIARA, P.O. Box 23-141, 
Taipei, 10617 Taiwan, R.O.C.}
\altaffiltext{3}{Northwestern University, Department of Physics and Astronomy, 2131 Tech Drive, Evanston, IL 60208, USA;
r-taam@northwestern.edu}

\begin{abstract}
Taking into account a range of parameters determined from the evolutionary
models and available observational data, the detailed
non-LTE spectrum for the primary star and  the irradiated
donor star in the AM CVn system SDSS J0926+3624 are constructed based
on the TLUSTY stellar atmosphere code. The combined spectrum of the
primary and the donor stars along with a multi-color blackbody spectrum
of the accretion disk that reproduces a detailed numerical model is
compared to the SDSS optical spectrum of the system. The 
photometric flux of the primary star inferred from eclipse observations is
compared with the synthetic spectrum. The model fit of the two
independent observations provides an upper limit on the distance of the
system for different effective temperatures of the primary.
In addition, an upper limit on the combined flux of the disk and the
donor in the infrared region wherein the contribution of the primary is
negligible is also determined.  It is shown that the spectrum of a
sufficiently cool donor can exhibit emission lines due to irradiation
from a hot primary and the emission features should be detectable
in the infrared even though the contribution of the flux from the disk
dominates. Thus, it is pointed out that infrared observations of the
system would provide important information
on the thermal state of the donor as well as useful insight on the 
thermal properties of the primary star and the accretion disk.  
\end{abstract}

\section{INTRODUCTION}

The only known eclipsing AM Canum Venaticorum object to date, SDSS J0926+3624 
\citep{and05}, belongs to the class of ultra compact binary systems
characterized by a pair of interacting white dwarfs with the shortest orbital
period among any other binary subclass.  Such a system is driven by the loss of 
orbital angular momentum by the emission of gravitational waves and is among a 
class of systems thought to constitute the largest fraction of interacting ultra
compact binaries in the Galaxy [see \cite{nel01,nel04}]. An understanding of their
properties is important for their detectability not only as individual 
candidate sources, but also as background sources of low frequency gravitational
waves [e.g.,\cite{nel04,sto05}]. 

A detailed understanding of the formation of SDSS J0926+3624 (hereafter J0926+3624) also allows one to
place important constraints on the efficiency of the common envelope phase of binary evolution in 
forming short period binary systems containing double white dwarf components [e.g., \cite{nel05};
for a review of the common envelope phase of evolution for the 
general binary population, see \cite{ibe93,taam00}]. For AM CVn systems, in general, 
their formation may involve a white dwarf or helium star channel 
\citep{nel01}, or an evolved hydrogen main sequence star channel \citep{pod03}. 
As shown by \cite{del07} their subsequent evolution is affected by the thermal 
state of the donor star.

The observational investigation of these double white dwarf systems provides important
insights into their nature and formation. For example, optical observations of the
line spectrum from the system carry the combined physical and chemical properties of
the accretor and the accretion disk while infrared observations of the
spectrum may provide important constraints on the thermal state, chemical composition, 
and surface gravity of the donor. 
As a result, atmospheric studies of the donor in combination with knowledge
of its distance can be used as an additional probe of the formation channel and 
evolutionary state of the system  \citep{roel07}. 

J0926+3624 offers a unique opportunity for probing the thermal properties of
the donor star since its surface gravity, essential for calculating the synthetic 
spectrum, is known.  In addition, the orbital separation between the accretor and
the donor, which determines the amount of irradiation onto the donor, is 
known from the component masses and orbital period. 
 
In this Letter, we present the model composite spectra
consisting of the non-LTE spectra of an irradiated donor, 
and a primary star as well as a multi-color blackbody
spectrum of the accretion disk. The model parameters
for the system are discussed in \S 2. The formalism for estimating the 
contribution of the accretion disk is also discussed in this
section. In \S 3, a brief description of the numerical methods 
employed for calculating the non-LTE theoretical spectra of both the
accretor and the donor stars is provided.  The results are described 
and discussed in \S 4 followed by our conclusions in the last section. 

\section{ATMOSPHERIC MODEL PARAMETERS}

The range for most of the model parameters of J0926+3624 is constrained 
by the modelling of eclipse light curves. Based on the theoretical work
of \cite{bil06} and the observational study of \cite{copper11} the mass
range of the primary (donor) is estimated as $0.79 - 0.89 \msun$ 
($0.027 - 0.038 \msun$). Taking the mass of the primary and the donor
stars as 0.84 $\msun$ and 0.029 $\msun$ respectively and the observed
orbital period of 28.3 min, \cite{del07} estimate the radius of the 
primary to be 0.01 $\rsun$ and the radius of the donor to 
be 0.043 $\rsun$. Hence, we set the surface gravity of the primary
at $\log(g)=8.3$ and that of the donor at $\log(g)=5.6$ where
the surface gravity g is in cm s$^{-2}$.
From the center-to-center distance and the radii of the components 
we find the surface-to-surface separation to be $1.67 \times 10^{10}\,cm.$ 
For irradiation from the photosphere of the primary star onto the 
surface of the donor star, the surface-to-surface separation is 
considered in our calculations. Although the incident radiation 
impinges on one face of the donor, we consider complete 
redistribution of the irradiated heat energy over the entire surface
of the donor.  As the donor star is likely to be tidally locked with
the primary star, its spin rotation period should be the same as
the orbital period. The small orbital period and the relative sizes of the 
components would tend to establish a complete 
redistribution of irradiated energy over the entire atmosphere of the donor star. 
The flux difference of the day and night sides
of Jupiter size exoplanets orbiting with a period of a few days
ranges from 10\% to 50\% \citep{bur08}. In the case of an
irradiated but self-luminous object with radius about an order
of magnitude less than Jupiter and orbiting with a period as 
small as 28 mins, the difference in the irradiated and
non-irradiated sides should be much less.   
 
The effective temperature of both the primary and
the donor remains uncertain owing to the absence of a direct
measurement of the distance to the system.  Recently,
\cite{copper11} estimated the contribution of the primary white
dwarf star from the eclipse depth, measured from multi-band, 
high time resolution light curves of the system. 
By fitting the flux in the ultraviolet and optical bands with
a synthetic spectrum they establish a lower limit to the
effective temperature of the primary $T_{\mathrm eff,P}$ as 17000K.
In previous work, \cite{del07} and \cite{bil06} theoretically estimated
$T_{\mathrm eff,P}$ to lie between 21000K and 24000K
based on its cooling and compressional luminosity. \cite{del07}
also predict the effective temperature of the donor star
$T_{\mathrm eff,D}$ to lie in the range between 1750K and 4500K.
Guided by the blackbody model of \cite{del07}, we adopt 
three different $T_{\mathrm eff,P}$ (18000K, 21000K and 25000K) and 
three different $T_{\mathrm eff,D}$ (3800K, 4000K and 4400K). 

The evolutionary model calculations by \cite{del07} estimate
the accretion rate to be about $10^{-10} \msun\,yr^{-1}$. 
There have been a few attempts to model the flux from the
accretion disk (see \cite{nag09} and references therein). 
However, all these models assume a steady state disk. Further,
from the variation in the location of a bright spot,
\cite{copper11} found that the geometry of the outer disk departs from
circular motion. A detailed non-LTE model of a non-steady state disk
with non-circular geometry is beyond the scope of the present 
investigation.  Therefore, we approximate the flux of
the disk by considering
\begin{eqnarray}
F_{disk} =\int^{R_{out}}_{R_{in}}2\pi r B(r)dr
\end{eqnarray}
where $B(r)$ is the Planck function corresponding to the radial
distribution of the effective temperature given by \cite{nag09} 
\begin{eqnarray}
T(r)=\left[\frac{3GM_P\dot{M}}{8\pi\sigma r^3}\left(1-\sqrt{\frac{
R_P}{r}}\right)\right]^{1/4}
\end{eqnarray}
where $M_P$ and $R_P$ are the mass and the radius of the primary respectively,
$\dot{M}$ is the mass accretion rate and $\sigma$ is Stefan-Boltzmann
constant. Following \cite{nag09} we take the inner radius
$R_{in}$ of the disk equal to $1.3 R_P$. The outer radius 
$R_{out}$ is taken to be $0.45a$ which is the maximum value
of the radius of the disk determined from the observation of
\cite{copper11} where $a$ is the orbital separation.
A composite blackbody spectrum reproduces
the non-LTE steady-state accretion disk 
model of \cite{nag09} well. The contribution of the disk to 
the total flux depends on the inclination angle and for the 
present object the inclination angle is as high as $83^o$
\citep{copper11}. A comparison of
the model spectra of \cite{nag09} for different inclination
angles implies that the flux should be reduced by a factor of 4 
to 7 when the inclination angle is $83^o$. Therefore, in order to 
obtain an upper limit on the contribution of the disk to the
total flux, we assume that at $83^{o}$ inclination angle the
flux is reduced by a factor of 4 as compared to its value at a face on view.

\section{NON-LTE MODEL ATMOSPHERES}  
  
The stellar-atmosphere code TLUSTY developed by Hubeny and 
colleagues \citep{hub88,hub95} and the related spectrum
synthesis code SYNSPEC \citep{hub94} are used for the investigation. In using
this code, we have adopted the atomic data and model atoms
provided on the TLUSTY homepage\footnote{http://nova.astro.umd.edu/index.html} \citep{hub07,hub03}.
The atomic spectral line database from CD ROM 23 of R. L.
Kurucz \citep{kur95} is used in SYNSPEC. The elemental abundances
adopted are listed in Table~1. The elements whose abundance is less than $10^{-5}$ 
are ignored as they do not significantly 
contribute to the total opacity. We have included 
micro-turbulence at the level of $2-5 \,kms^{-1}$ relevant for
high effective temperatures \citep{hub07}.
However, the overall spectrum is not altered even if this value
is increased by a few times.  For all the cases, the emergent
flux is integrated over all frequencies and the energy balance 
was checked for self consistency corresponding to the effective
temperature for models with and without irradiation.

In order to achieve convergent solutions for the atmospheric
model of the donor star, we have neglected the effects of 
convection. However, comparison with a model calculated by
using the atmosphere code PHOENIX [\cite{bar} 
and references therein] implies that convection
affects the spectrum in the optical where the flux of the donor
is negligible as compared to that of the primary or the disk. 

TLUSTY is well tested and widely used for constructing the 
spectrum of both hot stars \citep{hub07} and ultra-cool objects
such as brown dwarfs \citep{bur06}. We compared our results with a
set of similar models, but with different elemental abundances
constructed by using the atmosphere code PHOENIX. These models
were kindly provided to us by F. Allard and D. Homier (private
communication 2008).
  
\section{RESULTS}

Since the abundances of elements other than He are low, the
opacity of the atmosphere of the donor as well as the primary
is mainly determined by this constituent. Fig.~1 illustrates the thermal 
structure of the atmosphere of the donor star as a
function of pressure for different levels of irradiation from 
the accretor. It can be seen that at a pressure of about 100 
bar, the temperature increases rapidly. This rapid temperature
increase occurs as the Rosseland mean optical 
depth exceeds unity. The total pressure varies slowly with 
respect to the optical depth as the optical depth increases
from 1.2 at about 576 bar to 116 at about 667 bar
for $T_{\mathrm eff,D}=3800$K. When irradiation is included, the temperature varies 
significantly at optical depths less than unity.
For an effective temperature as low as 18000K for the
primary and as high as 4400K for the donor, a few
emission lines appear only in the infrared as can be seen from Fig.~2
(panel A). As the irradiation energy increases with the increase in 
the effective temperature of the accretor, the spectrum changes dramatically
with the appearance of emission lines in both the optical
and infrared wavelength regions (see panel B in Fig.~2). For a
fixed orbital separation, the emission features in the 
spectrum of the donor not only depend on the effective 
temperature of the accretor, but also on that of the donor
itself. As can be seen in panel B of Fig.~2, weak emission
lines are apparent from the optical, e.g., at wavelength
slightly greater than 0.5 $\mu$m  to infrared if the donor
is cooler ($T_{\mathrm eff,D}=3800$K) whereas strong emission lines 
are present in the entire spectrum if $T_{\mathrm eff,P}$ is increased to 21000K and to 
25000K even if the donor is hotter than 3800K.

The combined spectrum of the system is derived by multiplying the
surface area of each component with its flux and then 
summing them up with the flux of the disk. 
In Figs.~3 and 4 we present the synthetic spectra of the 
primary and the irradiated donor, the composite blackbody 
spectrum of the accretion disk as well as the combined flux
which is compared with the well calibrated and phase-averaged 
Sloan optical spectrum of the object. 
The observed spectrum from 0.38 to $0.92 \,\mu m$ is reproduced
from the SDSS archival data. The flux of the primary
at $\rm u'$, $\rm g'$ and $\rm r'$ bands are also presented for 
a comparison.  We have not performed any statistical analysis while 
comparing the synthetic spectra with the observed flux and our
fit is by eye. In Fig.~3, we also compare our synthetic spectra with that
constructed by using the atmosphere code PHOENIX. The overall agreement
over the emergent flux in the entire wavelength region adopted in our
investigation is considered as the evidence for a successful
implementation of the TLUSTY and the SYNSPEC codes.
   
Fig.~3 shows that for a model characterized by $T_{\mathrm eff,P}=
21000\,K$ and $T_{\mathrm eff,D}=4400 \,K$, the SDSS spectrum matches if the
distance of the system is about 495 pc and $\dot{M}=1.5\times10^{-10}
\,\msun yr^{-1}$. The object  J0926+3624
has another component, a spot. Unfortunately, the nature and
amount of flux contributed by the spot is not known. If the
contribution of the spot is comparable to that of the disk or
the primary, then the present fit of the SDSS data at 495 pc
implies an upper limit of the flux from the accretion disk.
On the other hand the observed flux of the primary is a few 
times less than the model flux of the primary. It should be
noted that the observed data of the primary is not calibrated
for extinction and therefore the actual flux may be slightly
higher. Therefore, the distance needed to fit both the observed
data may be considered as an upper limit. Figure~4 (panel A),
however, shows that the non-calibrated observed flux of the
primary fits well with a model characterized by 
$T_{\mathrm eff,P}=18000\,K$ if the distance is about 465 pc which is
consistent with the analysis of \cite{copper11}. At a distance 
d=465 pc, the combined flux also fits well with the SDSS data if $T_{\mathrm eff,D}=
3800\,K$. If the spot contributes significantly, then a decrease in the 
flux from the accretion disk is needed to fit the SDSS data at this
distance. It is worth mentioning here that the observation
of \cite{copper11} doe not indicate that the flux of the spot
is comparable to that of primary. The model characterized by
$T_{\mathrm eff,P}=25000\,K$ and $T_{\mathrm eff,D}=4000\,K$ is apparently ruled
out as the distance needed to fit the SDSS data is too high to
fit the observed flux of the primary at the three optical bands. A 
single disk model as described in \S 2 is considered in all the
above comparisons. 

For the range of the donor's effective temperature between 
3800K and 4400K and that of the primary between 18000K and
25000K, it is found that the flux of the donor dominates over
that of the primary at wavelength greater than about $1.1 \, 
\mu m$. However, the optimal flux of the accretion disk adopted in the
present simplified blackbody model is higher than that of
both the primary and the donor at all wavelengths. Therefore,
future observation at infrared wavelengths will enable an estimate of the 
combined flux of the accretion disk and the donor. Since the
mass and radius of the donor can be well constrained from the evolution
model, infrared observation would provide an excellent 
opportunity to estimate a range of the properties of the
accretion disk.

Further, an important finding in the present investigation is
the appearance of emission lines in the spectrum of the donor
due to the effect of irradiation.  Figs.~3 and 4 also show that
the combined flux does not dilute the emission feature 
originating from the donor's surface, even though the  
contribution from the disk is higher.
This, however is not obvious for any combination of the effective
temperature of the primary and the donor. 
As discussed earlier, the emission line strength becomes 
dominant when the temperature difference between the donor 
and the primary increases.  In the shorter wavelength region, 
atomic scattering  plays a dominant role over absorption.
Therefore, the emission lines occur in the near optical 
region only if the irradiation is sufficiently strong so 
that the scattering albedo is much less than 1.
However, the temperature-pressure profiles imply strong 
temperature inversion which indicates penetration of irradiation
energy into the atmosphere even below the region where the
optical depth is unity. In the region where the 
optical depth is comparable to unity, photons are absorbed and
re-emitted in the infrared. Therefore, even with relatively
weak irradiation, emission lines appear in the infrared. 
On the other hand, \cite{nag09} obtained few emission 
lines in the spectrum of the disk for an accretion rate higher
than $10^{-10}\,\msun yr^{-1}$. Therefore, detection
of emission lines in the infrared spectrum of the object would
provide important insight on the atmospheric properties of the
donor.  

\section{CONCLUSIONS}

Theoretical spectra were calculated for the two white
dwarf components in the AM CVn system J0926+3624 using the
non-LTE atmosphere code TLUSTY and SYNSPEC. A range of 
effective temperatures is adopted based on estimates derived
from theoretical binary evolutionary considerations for a
system formed in the white dwarf channel. Irradiation of the
cool donor star by the hot primary star is included by 
assuming complete redistribution of energy over the entire
atmosphere of the donor star. Observations indicate the presence of
a spot whose contribution to the total flux is unknown. In the
absence of a self-consistent non-steady state disk model,
a multi-color blackbody spectrum of the disk is considered and a 
maximum disk size is adopted as inferred from observation. Comparison
of the total flux derived by the present models places upper limits
on the distance of the object as well as the contribution of
the accretion disk to the total flux. The present findings emphasize 
that future infrared observations
of the system would provide a good estimation of the contribution
by the accretion disk to the total flux and hence important
insight on the thermal properties of the accretion disk. We also
predict emission lines in the infrared wavelength region, originating 
from the spectrum of the irradiated donor. The width of these
lines are dependent on the effective temperature of both
the donor and the accretor and hence high resolution spectrum 
may be needed to detect them.
If, however, the donor is sufficiently cool, as suggested by the
evolutionary models, the emission lines could be detected easily
providing an important diagnostic probe for the donor. 

\acknowledgements
We thank the referee for valuable comments and suggestions that
have improved the quality of this Letter.  Thanks are due to C. 
Deloye for providing the Sloan optical data of SDSS J0926+3624 and to 
F. Allard and D. Homier for providing a set of PHOENIX models for 
comparison. SS thanks G. Pandey for help in implementing the TLUSTY and 
SYNSPEC numerical programmes. RT acknowledges support, in part, from 
NSF grant no.  NSF AST-0703950 to Northwestern University.

\clearpage
\begin{figure}
\includegraphics[angle=0.0,scale=.70]{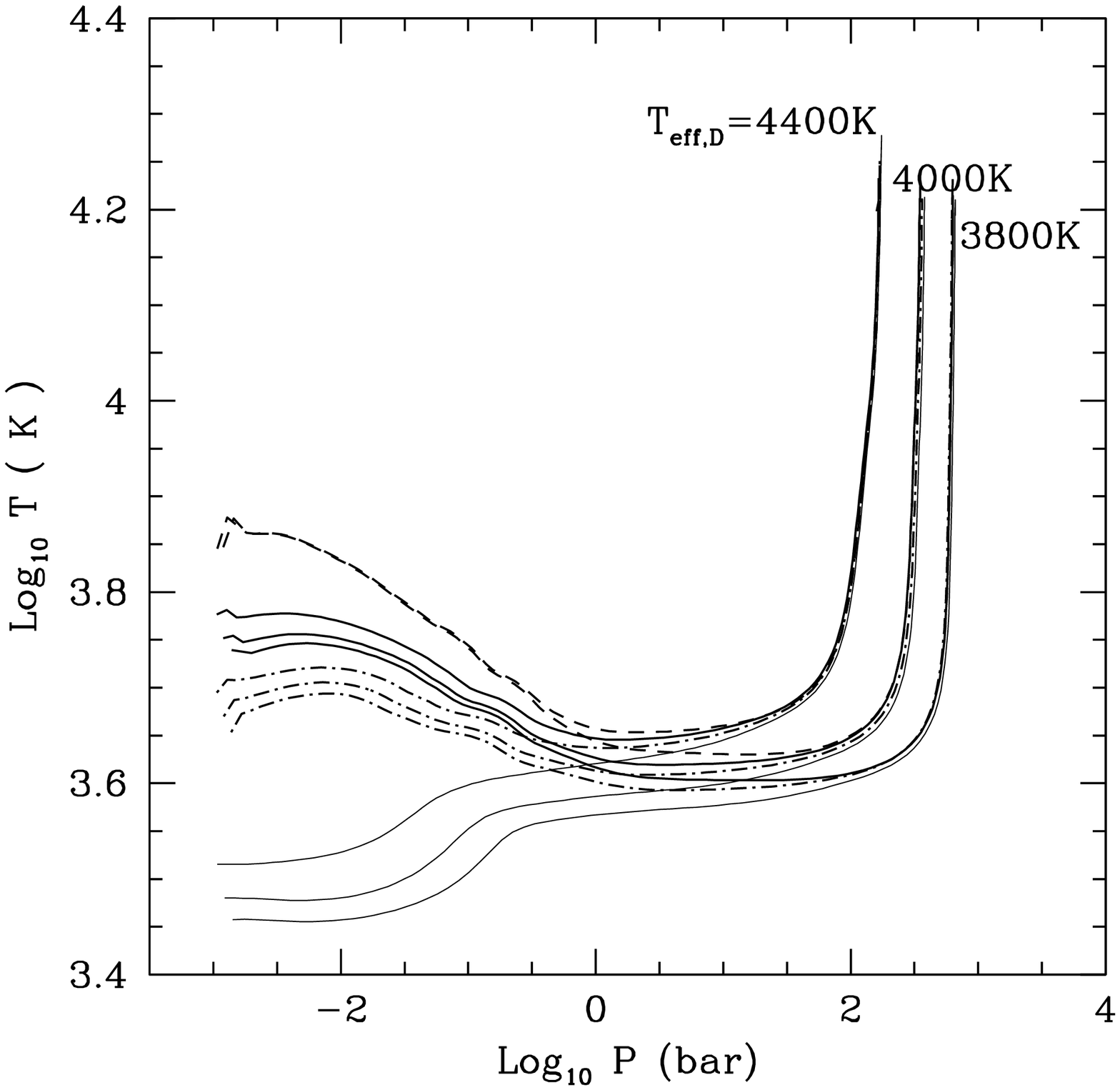}
\caption{ The Temperature-Pressure (T-P) profile of the donor
star. The numbers near the lines indicate the effective
temperature of the non-irradiated donor. The thin solid lines 
represent the T-P profiles of the non-irradiated donor while
the thick solid lines represent that with irradiation from a
primary with $T_{\mathrm eff,P}=21000\,K$ and the dashed line with 
$T_{\mathrm eff,P}=25000\,K$. The dash-dot lines represent the T-P 
profiles of the donor with irradiation from a primary of
 $T_{\mathrm eff,P}=18000\,K$.   
\label{fig1}}
\end{figure}

\begin{figure}
\includegraphics[angle=0.0,scale=.70]{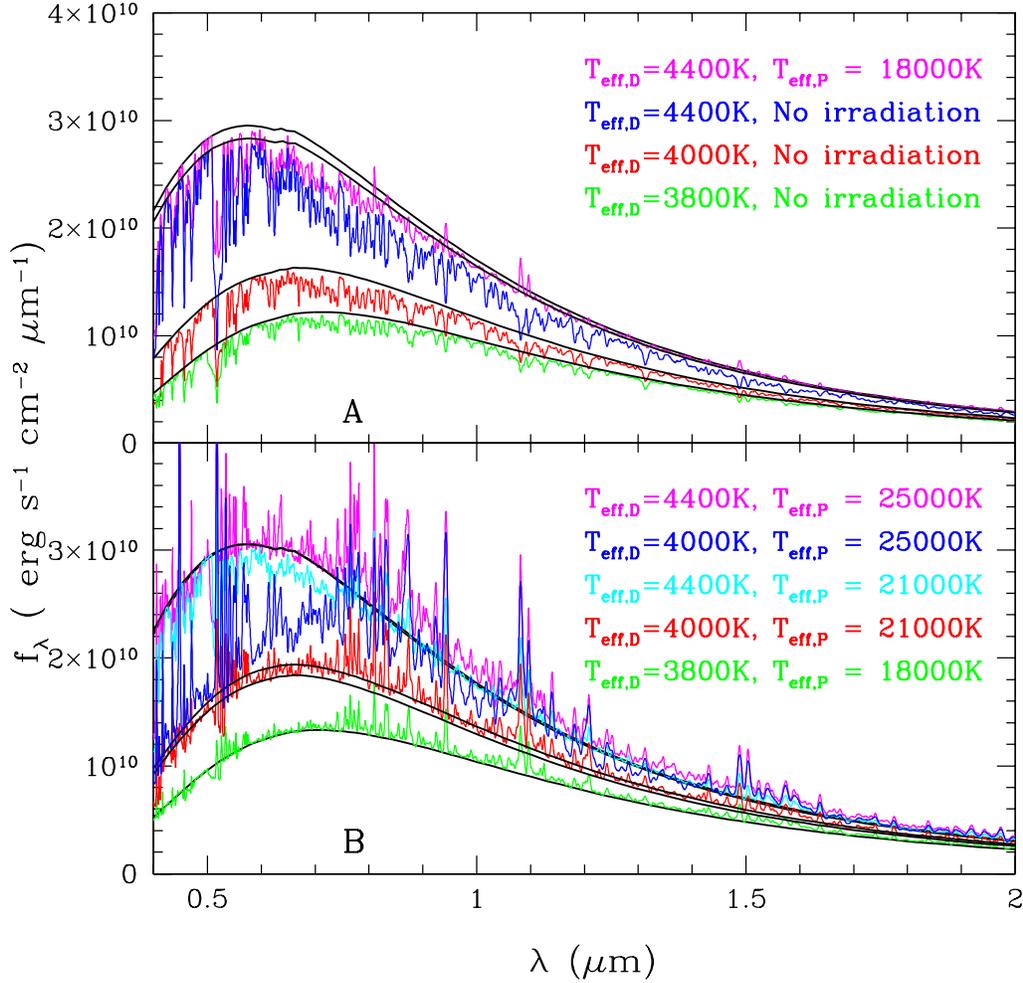}
\caption{The synthetic spectra of the non-irradiated (panel A)
and the irradiated (panel B) donor star. $T_{\mathrm eff,D}$ is the
effective temperature of the donor without irradiation and 
$T_{\mathrm eff,P}$ is that of the primary which irradiates the donor.
The black lines are the corresponding continuum flux.
The model with $T_{\mathrm eff,D}=4400\,K$, $T_{\mathrm eff,P}=18000\,K$ is also
included in panel A  for a comparison with the non-irradiated
case.
\label{fig2}}
\end{figure}

\begin{figure}
\includegraphics[angle=0.0,scale=.70]{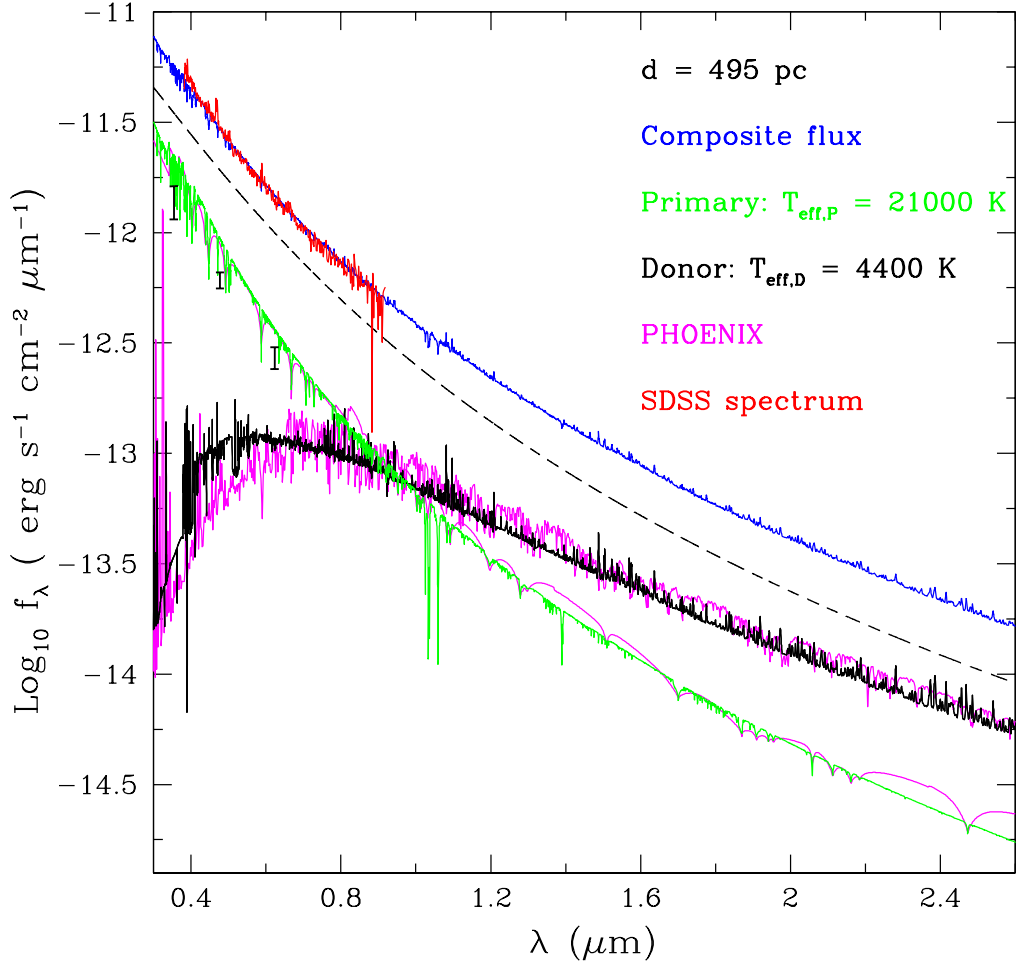}
\caption{Model fit of the Sloan optical spectrum (red lines) as
well as the photometric flux at u', g' and r' bands (black
error bars). The green line represents the spectrum of the 
primary with effective temperature $21000\,K$ and the black line 
represents that of the irradiated donor with $T_{\mathrm eff,D}=
4400\,K$. The dashed line represents the composite blackbody
spectrum of the accretion disk. The blue line is the combined
spectrum of the accretion disk, the primary and the donor stars. 
The results (with different elemental abundances) from the
atmosphere code PHOENIX are also presented for comparison.
\label{fig3}}
\end{figure}

\begin{figure}
\includegraphics[angle=0.0,scale=.70]{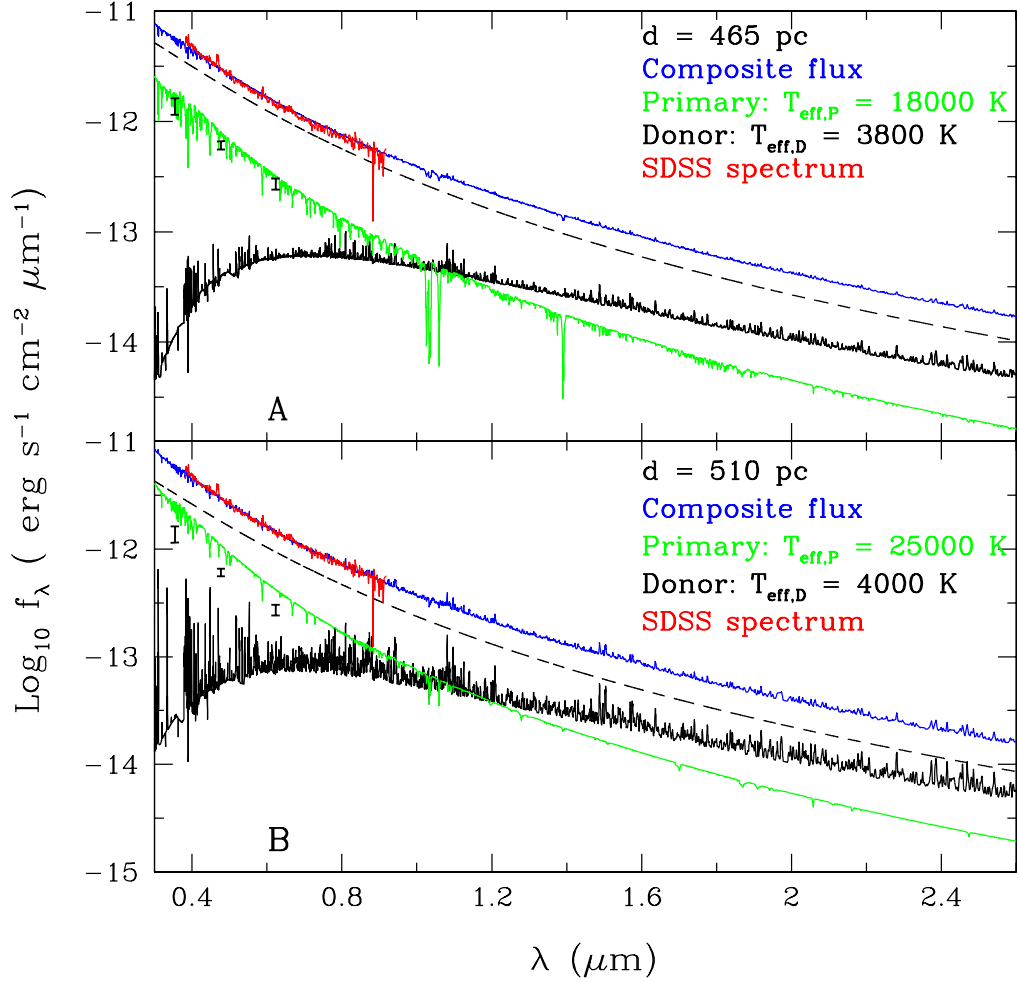}
\caption{Same as figure 3 but with different effective 
temperature of the primary and the donor stars.
\label{fig4}}
\end{figure}

\clearpage
\begin{table}
\begin{center}
\caption{The elemental abundances A  by number with respect to He adopted for the models. Comparison with solar abundances as derived by TLUSTY is presented in column 3.}
\begin{tabular}{ccc}
\tableline\tableline
Element &  A=N(element)/N(He) & A/A(Solar)    \\
\tableline 
He & $1.0$ & 10.0 \\
C & $4.79\times 10^{-5}$ & 0.145  \\
N & $3.55\times 10^{-3}$ & 42.7 \\
O & $2.71\times 10^{-4}$ & 0.401 \\
Ne & $3.83\times 10^{-4}$ & 3.19 \\
Mg & $1.21\times 10^{-4}$ & 3.18 \\
Si& $1.13\times 10^{-4}$  & 3.18 \\
S & $5.16\times 10^{-5}$ & 2.41  \\
Fe & $1.01 \times 10^{-4}$ & 3.2 \\
\tableline
\end{tabular}
\end{center}
\end{table}


\begin{thebibliography}{}
\bibitem[Anderson et al. (2005)]{and05} Anderson, S. F., et al.
2005, \aj, 130, 2230.
\bibitem[Barman, Hauschildt, Short \& Baron (2000)]{bar}
Barman, T. S., Hauschildt, P. H., Short, C. I., \& Baron, E.
2000, \apj, 537, 946.
\bibitem[Bildsten et al. (2006)]{bil06} Bildsten, L., Townsley,
D. M., Deloye, C. J., \& Nelemans, G. 2006, \apj, 640, 466. 
\bibitem[Burrows, Sudarsky \& Hubeny (2006)]{bur06} Burrows, A.,
Sudarsky, D., \& Hubeny, I. 2006, \apj, 640, 1063.
\bibitem[Burrows, Ibgui \& Hubeny (2008)]{bur08} Burrows, A., 
Ibgui, L., \& Hubeny, I. 2008, \apj, 682, 1277.
\bibitem [Copperwheat et al. (2011)]{copper11} Copperwheat, 
C. M. et al. 2011,MNRAS, 410, 1113.
\bibitem[Deloye et al. (2007)]{del07} Deloye, C. J., Taam, R. E.,
 Winisdoerffer, C. \& Chabrier, G. 2007, \mnras, 381,525.  
\bibitem[Hubeny (1988)]{hub88} Hubeny, I, 1988, Computer Physics Communications, 52, 103.
\bibitem[Hubeny, Hummer \& Lanz (1994)]{hub94} Hubeny, I.,
Hummer, D. G., \&  Lanz, T. 1994, A\& A 282, 151.
\bibitem[Hubeny \& Lanz (1995)]{hub95} Hubeny, I \& Lanz, T.
1995, \apj, 439, 875.
\bibitem[Iben \& Livio (1993)]{ibe93} Iben, I. Jr. \& Livio, M.
1993, PASP, 105, 1373.
\bibitem[Kurucz \& Bell (1995)]{kur95} Kurucz, R. L. \& B. Bell,
1995, Atomic Line Data, Kurucz CD-ROM No. 23. 
(Smithsonian Astrophysical Observatory : Cambridge, Mass.).
\bibitem[Lanz \& Hubeny (2003)]{hub03} Lanz, T. \& Hubeny, I. 
2003, \apjs, 146, 417. 
\bibitem[Lanz \& Hubeny (2007)]{hub07} Lanz, T. \& Hubeny, I.
2007, \apjs, 169, 83.
\bibitem[Nagel, Rauch \& Werner (2009)]{nag09} Nagel, T., Rauch,
T., \& Werner, K. 2009, A \& A, 499, 773.
\bibitem[Nelemans, Yungelson \& Portgies Zwart (2001)]{nel01}
Nelemans, G., Yungelson, L. R., \& Portegies Zwart, S. F. 2001,
A \& A, 375, 890.
\bibitem[Nelemans, Yungelson \& Portgies Zwart (2004)]{nel04}
Nelemans, G., Yungelson, L. R., \& Portegies Zwart, S. F. 2004,
MNRAS, 349, 181.
\bibitem[Nelemans \& Tout (2005)]{nel05} Nelemans, G., \&
Tout, C. A. 2005, MNRAS, 356, 753.
\bibitem[Podsiadlowski, Han \& Rappaport (2003)]{pod03} 
Podsiadlowski, P., Han, Z., \& Rappaport, S. 2003, MNRAS, 340,
1214.
\bibitem[Roelofs et al. (2007)]{roel07} Roelofs, G. H. A. et al.
 2007, ApJ, 666, 1174.
\bibitem[Stroeer, Vecchio \& Nelemans (2005)]{sto05}
Stroeer, A., Vecchio, A., \& Nelemans, G. 2005, \apj, 633, L33.
\bibitem[Taam \& Sandquist (2000)]{taam00} Taam, R. E. \&
Sandquist, E. L. 2000, ARAA, 38, 606.
\end{thebibliography}
\end{document}